\documentclass[final,5p,times]{elsarticle}

\usepackage{graphicx}
\usepackage{amssymb}
\usepackage{colortbl}
\journal{Nuclear Physics B}

\begin{document}
\begin{frontmatter}

\title{CTA - A Project for a New Generation of Cherenkov Telescopes}
\author{Michele Doro\corref{cor1}}
\address{University of Padova \& INFN Padova, v. Marzolo 8, 35131
  Padova, Italy}
\ead{michele.doro@pd.infn.it}
\cortext[cor1]{for the CTA consortium.}

\begin{abstract}
Gamma-rays provide a powerful insight into the non-thermal universe
and perhaps a unique probe for new physics beyond the standard
model. Current experiments are already giving results in the physics
of acceleration of cosmic rays in supernova remnants, pulsar and
active galactic nuclei with a hundred sources detected at
very-high-energies so far. Despite its relatively recent appearance, very high-energy gamma-ray astronomy
has proven to have reached a mature technology with
fast assembling, relatively cheap and reliable telescopes.  The goal
of future installation is to increase the sensitivity by a factor ten
compared to current installations, and enlarge the energy domain from
few tens of GeV to a hundred TeV. Gamma-ray spectra of astrophysical origin
are rather soft thus hardly one single size telescope can cover more
than 1.5 decades in energy, therefore an array of telescopes of $2-3$
different sizes is required. Hereafter, we present design
considerations for a Cherenkov Telescope Array (CTA), a project for a
new generation of highly automated telescopes for gamma-ray
astronomy. The status of the project, technical solutions and an
insight in the involved physics will be presented. 
\end{abstract}

\end{frontmatter}

\section{Introduction}\label{sec:intro}

\subsection{Gamma-ray astronomy as a probe for cosmic rays.}

\noindent
Despite its origin goes back to only few decades ago, ground-based
gamma-ray astronomy has already demonstrated to be a mature scientific
technique to probe non-thermal phenomena in the universe,
where cosmic ray (CR) particles are accelerated to extremely high
energies. CRs of non-thermal origin can be  accelerated
either directly at the place of origin, for example nearby the surface of
a 
fast rotating pulsar, or may gain energy in cosmological times through
interaction with irregular cosmic magnetic fields or shock wave
fronts. CRs  cover a vast range of energies, from $10^9$ to
$10^{21}$~eV. Below 100~GeV the CR interaction with the solar wind is
efficient and they are absorbed. Their flux covers more than 32 
orders of magnitude, and they are distributed roughly according to a
power-law of spectral index $-3$. There are many publications about
CRs, but still a conclusive evidence about how and where they are
accelerated, is missing~\cite{Ginzburg:1964}.  
CRs are mainly composed of protons and helium, with
smaller percentages of heavier elements and electrons. Due to the
rapidly falling fluxes of CRs, it is experimentally difficult to cover an energy
band larger than few decades in energy by a single
experiment. Therefore different experimental techniques  are needed for
different energy ranges. They are
typically investigated with balloon-borne calorimeters and ground-based
detectors. For a recent review, see, e.g., Ref.~\cite{Aharonian:2008zz}.

After the second half of last century, it was recognized that
gamma-rays (GRs) could provide a useful tool to investigate the origin
of CRs. First of all, GRs are neutral, thus they
travel undeflected by inter-stellar or inter-galactic magnetic fields
and therefore trace back their origin, and second, they interact with
local radiation/dust fields thus providing useful information on the
morphology of the emission region. 

\subsection{Scientific targets}
As in the case of CRs, it is not possible to construct a single
observatory which could cover the entire GR spectrum because of the
rapidly falling flux with GR energies.
 From an
experimental point of view, two sub-ranges are defined at high energy
(HE; MeV--GeV) and very-high energy (VHE; GeV-TeV). For many sources
the combined power emitted at these energies overcome the total power
emitted at other wavelengths, and therefore HE and VHE astrophysics
has attained large attention over the last
years~\cite{Aharonian:2008zz}.  

The goals of GR astronomy could be divided in three categories: 
the \emph{galactic targets}, like pulsar and pulsar-wind
nebulae (PWN), supernova remnants (SNR) and star-active region like
OB-associations or binary systems. Another fundamental target in this
family is the galaxy center (GC). Among the \emph{extragalactic
  targets}, stand the active galactic nuclei (AGN), particularly
blazars and radio-galaxies. Galaxy clusters, starburst and merging
galaxies are also interesting targets. Finally, 
gamma-ray bursts (GRBs) are also among this category. There is a third family of
observation about \emph{fundamental physics}, which can be studied
within GR astronomy. At first place is the study of
dark matter (DM) annihilation or decays, which could have very clear
GR signatures. The extragalactic background light (EBL) can be
studies through the GR absorption from distant targets to
understand the universe transparency. Other questions like axion
physics, Lorentz invariance violation, antiparticle asymmetry can also be
addressed. For a recent review, see, e.g., Ref.~\cite{Buckley:2008}.

\subsection{The IACT technique}
Very high energy GRs impinging on the earth, interact with
atmospheric nuclei and generate an electromagnetic shower. The showers extend over
several kilometers in length and few tens to hundreds of meters in width, and their
maximum is located at $8-12$~km altitude, in case of vertical
incidence. For gammas below around 100 TeV, the shower particles stop 
high up in the atmosphere, and can not be directly detected at ground. However,
a sizeable fraction of the charged secondary shower particles, mostly electrons and
positrons in the shower core, move with ultra-relativistic speed and emit Cherenkov
light. This radiation is mainly concentrated in the near UV and optical band and
therefore passes mostly unattenuated to the ground, with minor losses due to Rayleigh
and Mie scattering and Ozone absorption. Imaging Atmospheric
Cherenkov Telescopes (IACTs) reflect the Cherenkov light
at the focal plane where a multi-pixel camera
records the shower image. The technique was pioneered by the Whipple
experiment, which first detected the Crab Nebula at VHE, in 1989.

\section{Current IACTs. Technologies and Selected Results}

\subsection{Current experiments}

\begin{figure*}[ht!]
\centering
\includegraphics[width=0.7\linewidth]{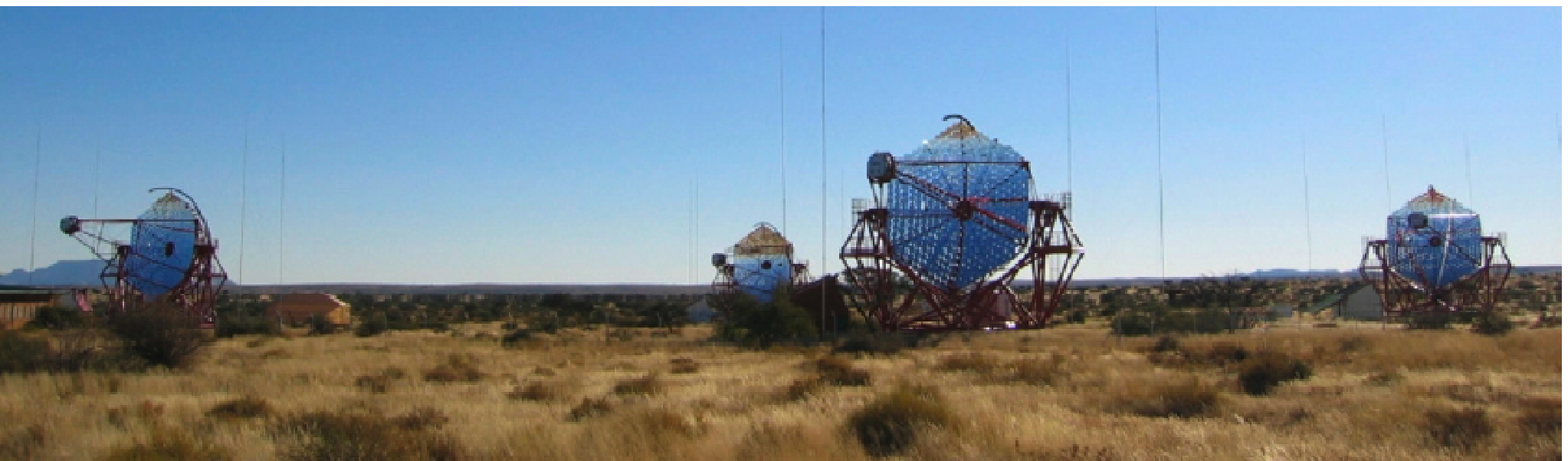}
\includegraphics[width=0.7\linewidth]{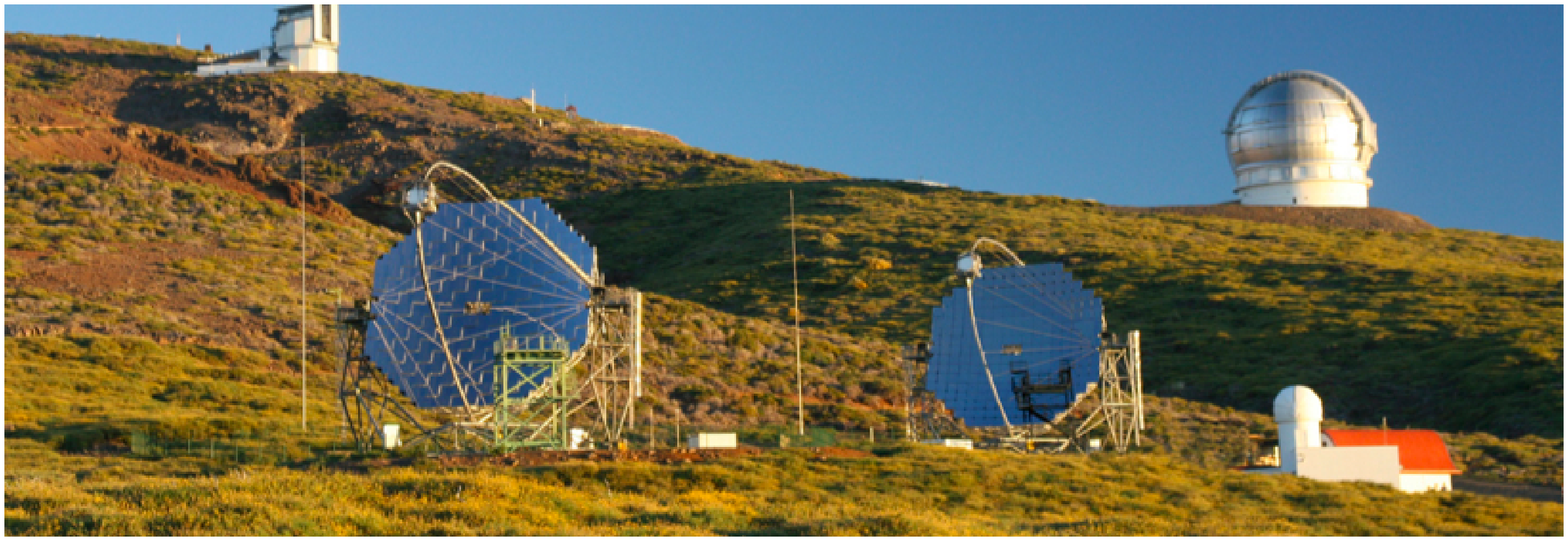}
\includegraphics[width=0.7\linewidth]{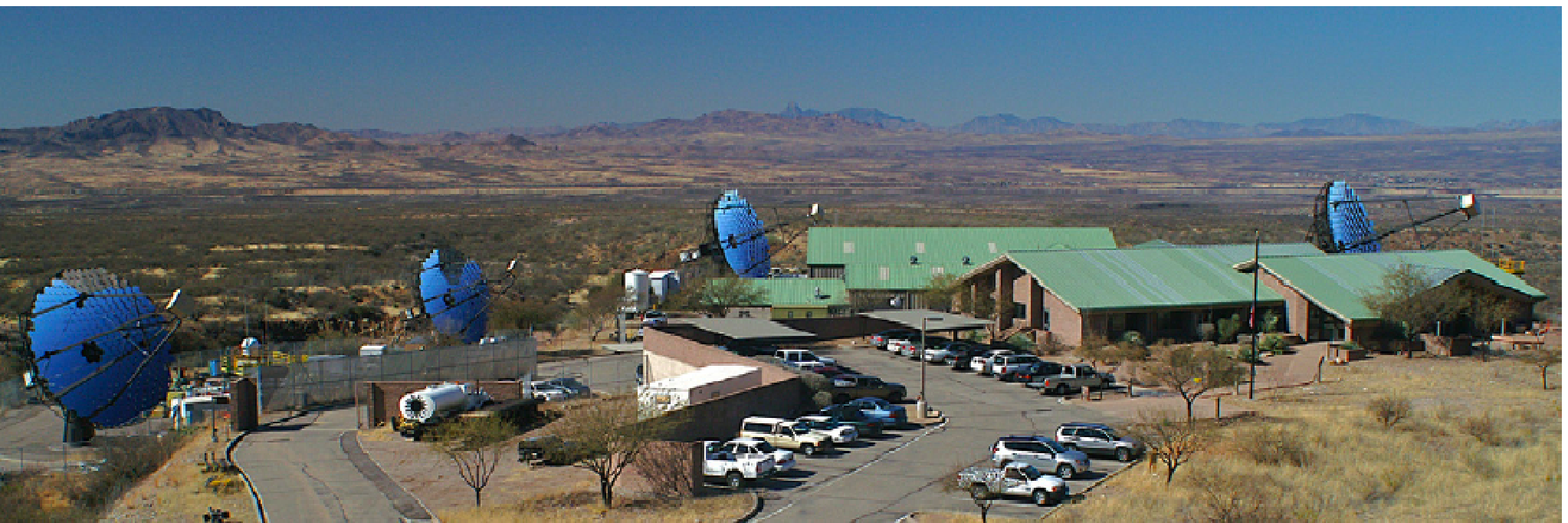}
\caption{\label{fig:iact}From top to bottom, the HESS array of four
telescopes located in Namibia; the MAGIC array of two 17~m diameter
telescopes located in the Canary island  La Palma; the recently
completed four IACT array VERITAS at Mt Hopkins, Arizona.} 
\end{figure*}

Currently, the world largest ground-based IACTs are HESS, MAGIC and
VERITAS\footnote{www.mpi-hd.mpg.de/HESS, wwwmagic.mppmu.mpg.de,
  veritas.sao.arizona.edu} (Figure~\ref{fig:iact}). HESS is an array of 4 clone telescopes,
each of 12~m diameter, located in the Gamsberg mountain in Namibia and
is operating since 2003 with a very high scientific impact. A fifth telescope, dubbed HESS-II, of 28~m
diameter, is under construction at the center of the array, and its
completion is foreseen for 2010. MAGIC has operated since 2004 with a
single dish of 17~m diameter and parabolic profile in the Canary island La
Palma in Spain. Despite the use of a single reflector does not
guarantee the sensitivity of an array, its world-largest dish allowed
to reach  the lowest energy threshold of the IACT
technique, performing for the first time observation below 
100~GeV with this technique. Recently MAGIC has
detected the first ever observed GR pulses at 25~GeV from the
Crab pulsar~\cite{Aliu:2008}. Recently this
year, a second clone MAGIC telescope, dubbed MAGIC~II, has been
inaugurated. The use of the stereoscopic system will allow MAGIC
phase-II to reach the sensitivity of larger arrays. A more recent
experiment was started in the Arizona desert in USA, following the
successful experience of the Whipple experiment. VERITAS has soon reached
the expected performance, with a sensitivity comparable to HESS and is
starting to collect important scientific results.

\subsection{IACT, a well proven technology.}
Despite its origin traces back to the late 70's, one
can allege that IACT has already reached a mature
technological development. In the following, the basic features will
be discussed. \emph{Mounting.} HESS, MAGIC and VERITAS have an
alt-azimu\-th mount, but while the former two have structures rotating on
wheels, VERITAS has a sole central mast and therefore an easier design.  HESS is built from stainless steel while MAGIC is
made of lighter carbon-fiber reinforced-plastic (CFRP) tubes. \emph{Mirrors.} The demands 
for the quality of mirror facets for IACTs are quite less challenging than
for optical telescopes, because IACTs focus on the intrinsic-aberrated
Cherenkov light from atmospheric showers. On the other hand, the
facets are exposed to the environment and must be robust. A
classical solution is the use of quartz-coated aluminized glass
substrate. An innovative solution was used by MAGIC, based on
all-aluminum diamond-milled sandwiches with hexcell honeycomb
interspaced. This technique has proven to produce mirrors with
reflectivity loss more than 5 times smaller compared to standard
techniques. MAGIC~II is also featuring novel glass-aluminum sandwich
mirror facets, whose performance is under investigation and with
interesting prospects for the future. \emph{Focal-plane instrumentation.} HESS,
MAGIC and VERITAS have a focal-plane instrumentation composed of a
multi-pixel camera of photomultiplier tube (PMTs; respectively
$960,576,1038,499$ for HESS, MAGIC-I, MA\-GIC-II and VERITAS). PMTs are an optimal solution due to their high gain and fast
read-out. A drawback is their limited photon conversion efficiency for
Cherenkov photons, currently about 20-25\%. More performing devices
are under research, as will be discussed later. \emph{Read-out
  electronics.} The short duration of Cherenkov signals demands at
least a MHz sampling capacity and a signal forming, whereas GHz
sampling is already in use in MAGIC and HESS. Such technology is continuously
evolving and research is ongoing. \emph{Trigger system.} The trigger
system is multifold: the basic discrimination is at the photoelectrons
level to adjust to different illumination conditions (moon, galactic
or extragalactic targets); a second-level trigger is typically
topological when a cluster of PMTs are activated. This allows to
reject spurious events from the light of the night sky. Other
triggers are used to synchronize telescopes in the array or make more
advance logical trigger. \emph{Calibration.} Despite an exact energy
calibration is impossible for IACT because of a missing calibrated
GR source in the sky\footnote{This fact will be strongly
  mitigated once IACT data could be cross-calibrated with satellite
  GR detector data (e.q Fermi).} and because of the varying
conditions of the atmosphere, several instrumental calibration methods are
normally applied. In the future, particular interest will be focussed on
improving devices and methods for calibration to reduce systematic
biases in particular regarding the  energy resolution.

\subsection{Selection of scientific results}
Interesting reviews of scientific results of GR astrophysics can be
found in Refs.~\cite{Aharonian:2008zz,Buckley:2008}.

\section{Towards a precision gamma-ray astronomy. \\Physics motivation
  for CTA.}
Despite the promising achievements from the current generation of
IACTs, there is a number of limitations that the future generation
will overcome: $a)$ IACTs of current generation are sensitive in a limited energy range, from 100~GeV
  to 50~TeV. At the lower end, IACTs are limited by the background
  from atmospheric hadronic showers. At the high end, the limit is
  posed by insufficient statistics; $b)$ Due to the lack of a calibrated cosmic GR source, IACT
  spectral reconstruction is limited by systematic bias and statistical
  uncertainties on the energy reconstruction; $c)$ They have a limited aperture, with typical field of view
  (FOV) of the order of $3-5^\circ$~diameter; $d)$ They have a limited angular resolution which currently states
  around few arcmin; $e)$ They have limited collection area; $f)$ They
  are rather poorly automatized.

On the other hand, from a physical point of view, there are strong
arguments to improve in the following aspects:

\begin{enumerate}
\item Decrease the energy threshold to few tens of GeV
\item Acquire sensitivity beyond 50~TeV
\item Increase sensitivity in the bulk range 100~GeV, 50~TeV
\item Improve energy and angular resolution
\end{enumerate}

\begin{figure}[!]
\centering
\includegraphics[width=0.98\linewidth]{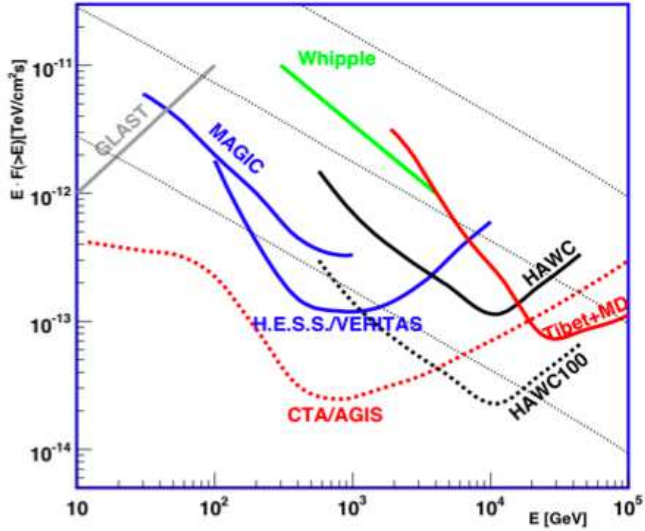}
\caption{\label{fig:cta_sens}Integral sensitivity for a Crab-like
  spectrum for several current IACT and expected for CTA/AGIS
  (5$\sigma$, 50~h)
  and Fermi/GLAST (5$\sigma$, 1~yr).} 
\end{figure}

The accomplishment of the above goals is intimately related to the
overcome of the technological questions discussed at the beginning of
this section. For this reason, a new generation of IACT is under
design now with expected performance well above the current
generation, as shown in Figure~\ref{fig:cta_sens}. We believe that with CTA we are heading toward
a new era of \emph{precision GR astronomy}. 
In the following, the above arguments are discussed in more detail.

\subsection{Low-energy physics (sub-50 GeV)}
There is a lot of expectations in the GR physics below 50~GeV. MAGIC has opened the field of sub-100~GeV GR
astronomy, although this is currently done only with poor significance. Observations with larger sensitivity in this region will
have several important consequences for galactic and extragalactic
physics, and maybe for fundamental physics. The expectations in term
of sensitivity are shown in Figure~\ref{fig:cta_sens}.

Pulsars are compact objects, residual of exploded massive stars, with
extreme physical conditions in their proximity: high magnetic field
($10^{12}-10^{15}$~Gauss) and strong particle and radiation
winds. They are observable at all wavelengths, particularly in the
radio band, since many years, with
a catalog of thousand objects. On the other hand, only around
60~pulsars were recently observed by Fermi at HE GRs~\cite{Abdo:2009mg}, and only
the Crab pulsar was observed by MAGIC at VHE
energies~\cite{Aliu:2008}. This is related to the fact that a cutoff in the
emission is expected in the region between few GeV and few tens of
GeV. By studying this energy band, CTA will provide the final answer
to the acceleration mechanism in pulsar, because the
exact location of the cutoff strongly depend on the pulsar
acceleration model (e.g., higher energy cutoffs foresee acceleration far
from the pulsar surface) and the high energy tail of the Fermi spectra
will be completed and cross-checked with higher significance. 

Farther from the
pulsar core, the pulsar particle wind is typically able to sweep up the
surrounding medium through shock mechanism and forming the
PWN. Ultrarelativistic electrons inside the nebula emits synchrotron
GRs in the HE and VHE regime as a result of the interaction with local
magnetic fields. The exact location of the synchrotron peak is still
unknown and may depends on the exact feature of the emitter. Low-energy
studies will permit to fully characterize this important radiation
loss mechanisms in the nebula and the interaction with the surrounding
medium. 

Even farther away from the remaining of a collapsing supernova
star, one can find the remnants of the material ejected at the
explosion. These are called SNRs. They are expanding clouds which
propagate at fast velocity. All 
the charged particles inside the clouds are heated up and accelerated
by statistical mechanisms described by the Diffusive Shock
Acceleration (DSA) model~\cite{Drury:1983} based on the second Fermi
acceleration mechanism. Those models predict acceleration of
electrons and/or light hadrons. While the electronic acceleration has
several evidences, there is not yet a clear indication that hadronic
acceleration takes place at SNRs. A clearer view will come from low-energy
studies because the predictions for the two emissions differ
substantially at these energies. 

The same argument for acceleration mechanism applies in the case of
AGNs, particularly in the case of the steep spectrum blazars. 
Finally, the importance of low-energy studies should not be
underestimated, to complete the Fermi catalog of HE emitters in the VHE tail
with possibly a larger significance than what Fermi can obtain. 
 This
will play a fundamental role in the modelization of the source
emission, and will provide a unique way to understand the nature of
the tens of yet unidentified Fermi sources~\cite{Abdo:2009mg}. 

\subsection{High-energy physics (above 50~TeV)}
It is strongly believed that typical GR spectra of astrophysical
source have a bimodal distribution with one peak at lower energies to
synchrotron emission from charged particles, and a secondary peak at
higher energies due to inverse Compton scattering of VHE electrons on
seed IR photons. In case of galactic objects, at VHE one may expect to
observe power-law GR spectra with cutoffs due to intrinsic
mechanisms. On the other hand, GRs from distant blazars suffer a severe
attenuation after pair production with local IR-UV photons of the
extragalactic background light (EBL). In all current IACT data, the
evidences for spectral cutoffs both for extragalactic and galactic
objects are rather poor. There is no simple justification for this
evidence. It is possible however that cutoffs are placed at higher
energies. 

CTA will explore this region with unprecedented
significance. This will allow to understand the
acceleration mechanism in galactic objects like SNRs, and
discriminating the hadronic vs leptonic models. Hadrons are expected to produce higher energy GR
due to the faster electron energy loss. Additional interesting
targets will be understood in this domain. For example, the nature of
the ultrarelativistic jets of micro-quasars and of binary
objects with massive infall of materials will be clarified.

Finally, detecting eventual flares from distant object like AGNs at
super-TeV energies, would allow to increase the prospects for a
detection of Lorentz invariance violation.

\subsection{TeV-regime (0.1-50~~TeV). Bulk research field.}
The greater effort for CTA will be concentrated in boosting the
instrumental sensitivity of at least a factor 10 in the actual
sensitive region of current IACTs, i.e. between 100~GeV and
50~TeV, reaching a level of $10^{-3}$~C.U.\footnote{1~C.U. (Crab Unit) $=
  1.5\times10^3$~(E/GeV)$^{2.58}$~ph~cm$^{-2}$~s$^{-1}$~TeV$^{-1}$}
sensitivity in this range (see Figure~\ref{fig:cta_sens}). This will promote the GR astrophysics
science to the level of \emph{GR astronomy}. In fact, for the first
time, a full VHE sky coverage will be obtained, with a thousand of new
VHE~GR sources expected. In parallel, the first VHE~GR catalog of
emitters will be released, making a big step toward a complete
understanding of the non-thermal emission scenario. Entire classes of
sources could be understood once their acceleration mechanisms will be
revealed. 

The increase in sensitivity will in fact not only have the consequence
of requiring shorter observation time. An increase in sensitivity will
allow to: $a)$ increase the possibility for follow-up observation and
finer resolution time-variability, and $b)$ improve the resolution for
morphological observation. Actual telescopes are at best sensitive
enough  to detect variations on the time scale of minutes. By increasing the sensitivity
of a factor 10, a sub-minute resolution will open the door to
understand the complex phenomena of GR flares, directly
connected to the acceleration mechanisms and the local
environment. Also of utmost importance is the morphological studies of
extended GR emission from galactic sources like SNRs. In this case, a
uniform sensitivity must be guaranteed over the entire FOV of the
instruments and therefore an overall gain 
in sensitivity must be accomplished. Morphological studies allow a
preciser comparison with other wavelengths and a clearer view of the local
interactions. 


\section{General technical ideas on CTA}
\begin{figure*}[!ht]
\centering
\includegraphics[width=0.9\linewidth]{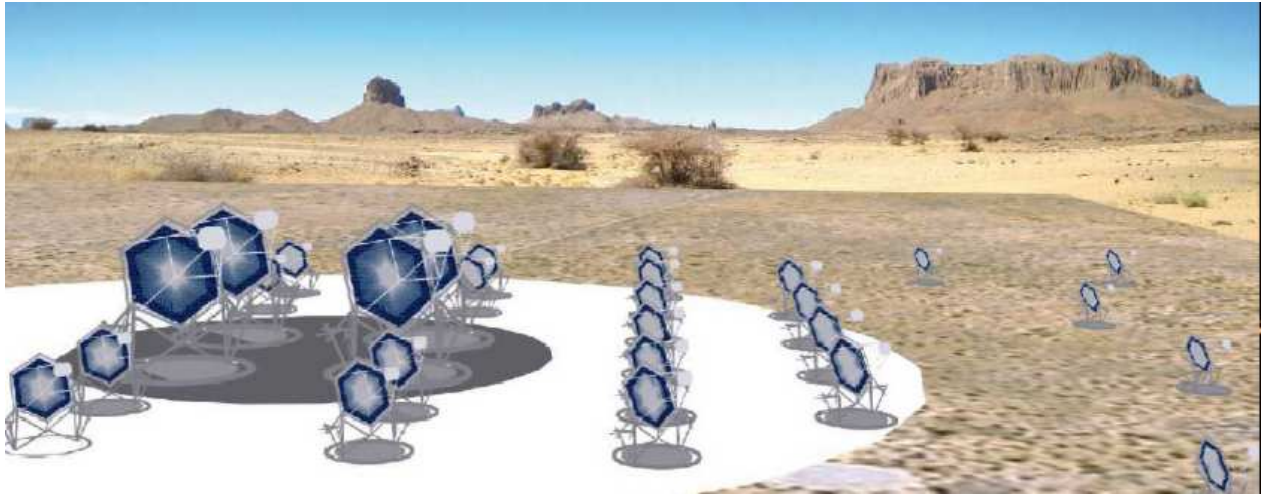}
\caption{\label{fig:cta_artist}Artistic view of the compound different
size telescopes CTA system. The area coverage is of $1-10$~km$^2$.}
\end{figure*}

To accomplish with the physical requirements specified in the previous
section and to maintain an overall high technical performance, the CTA
concept is based on few general ideas:

\begin{enumerate}
\item Increase the array from 4 to around 100 telescopes;
\item Distribute them over a large area ($1-10$~km$^2$);
\item Make use of telescopes of $2-3$ different sizes;
\item Take advantage of  well-proven technology of current IACTs;
\item High automatization and remote operation;
\item Run array as observatory and open to astronomer community.
\end{enumerate} 

\subsection{Basic array design}
The design of the array is artistically depicted in
Figure~\ref{fig:cta_artist}. The CTA
project is being designed both to provide an
expansion of the energy range down to a few tens of GeV and up to about 100 TeV
and with at least 10 times improvement in
sensitivity compared to current installations. This can only be
achieved by combining $a)$ many telescopes distributed over a large
area of at least 1~km$^2$ and $b)$ using telescopes of different
sizes. CTA is planned to comprise 
about a hundred telescopes of $2-3$ different sizes: several small
size telescopes (SST) of 6~m diameter, several medium size telescopes
(MST) of 12~m and few large size telescopes (LST) of 23~m
diameter. However, the
number of the telescopes, their size, their configuration and the overall
performance are still under investigation and the final layout will
come out after Monte Carlo optimization.

Few LSTs should catch the sub-100 GeV photons
thanks to their large reflective area. To maintain the time stamp of
the showers, they will have probably a parabolic shape. To avoid the
intrinsic optical aberrations due to this profile, LSTs will probably
have limited size FOV ($3-4^\circ$) while large telescope $f/D>1.2$ ratio
presents the technical challenge of displacing the camera at more than
28~m from the reflector. Technologically, the LST will be the most
challenging telescope. A design is current under development.

Several tens of MSTs will perform the bulk TeV search. Those telescopes will come from the well-proven
experience of HESS and MAGIC collaborations. The main goal is to
reduce the costs and maintenance activities. They will constitute the
core of the array, and will perform the fundamental task of vetoing
the LST triggers to reduce the hadronic background. Several different
designs are currently taken into account and the construction of the
first prototypes is expected in few years from now.

Finally, several tens of SSTs will complete the array
to perform the super-TeV search. They will be very simple in
construction, contributing to a small percentage of costs of the full
array, and distributed in between and around the core array of MSTs. 

\subsection{Improved angular and energy resolution}
The improvement of energy and angular resolution is also a very
important point. The angular resolution should be kept as
low as possible in the entire FOV. Current theoretical limits are
discussed in~\cite{Hofmann:2006wf}. This has the important impact of
a) Avoiding the source confusion; b) Contributing in the
discrimination between acceleration of hadrons and leptons. They have
different free-streaming lengths and the GR emission is strongly 
shaped by local interactions; c) Possibility to better cope with
data at other wavelengths. All this will give key information to
discriminate acceleration mechanisms in SNRs, PWNs, binaries,
micro-quasar and the GC. An improved angular resolution
can be obtained with performing optical system and by keeping optical
aberrations low.

An increase in energy resolution (down to 10\%) has a direct
consequence in boosting the capability to observe cutoffs such those expected
in pulsars and EBL-absorbed AGNs. Cutoffs are also expected in GR
spectra from DM candidates at the DM mass. Finally, a better
energy resolution can provide higher quality measurement of Lorentz
invariance. This can be achieved by an overall improvement both in the
calibration of the detector (light sensor, photon conversion, etc.)
and in a deeper monitoring of the atmosphere through transparency
instrumentation, like LIDARs.

\subsection{Subsystems}
From the technological point of view, the major effort is currently
concentrated on the working packages responsible for telescope design,
mirror facets development, electronic and focal-plane
instrumentation. Different projects for SSTs, MSTs and LSTs are
under design at several institutes to stimulate competition
and technological development. Mirrors constitute an important
challenge because they contribute to a sizeable part of the costs. A
major effort is ongoing to build fast assembling, reliable mirrors
with reduced optical degradation with time. A review of the different
techniques under study can be found in Ref.~\cite{Doro:2009}. The
focal-plane instrumentation also represents a challenge in technology
and cost. The current baseline design envisages the use of high
conversion efficiency PMTs. Studies are ongoing also on GaAsP hybrid
photon detectors (HPD)  and Geiger avalanche photon detectors (GAPD).

\subsection{Operation modes}
\begin{figure}[t!h]
\centering
\includegraphics[width=0.32\linewidth]{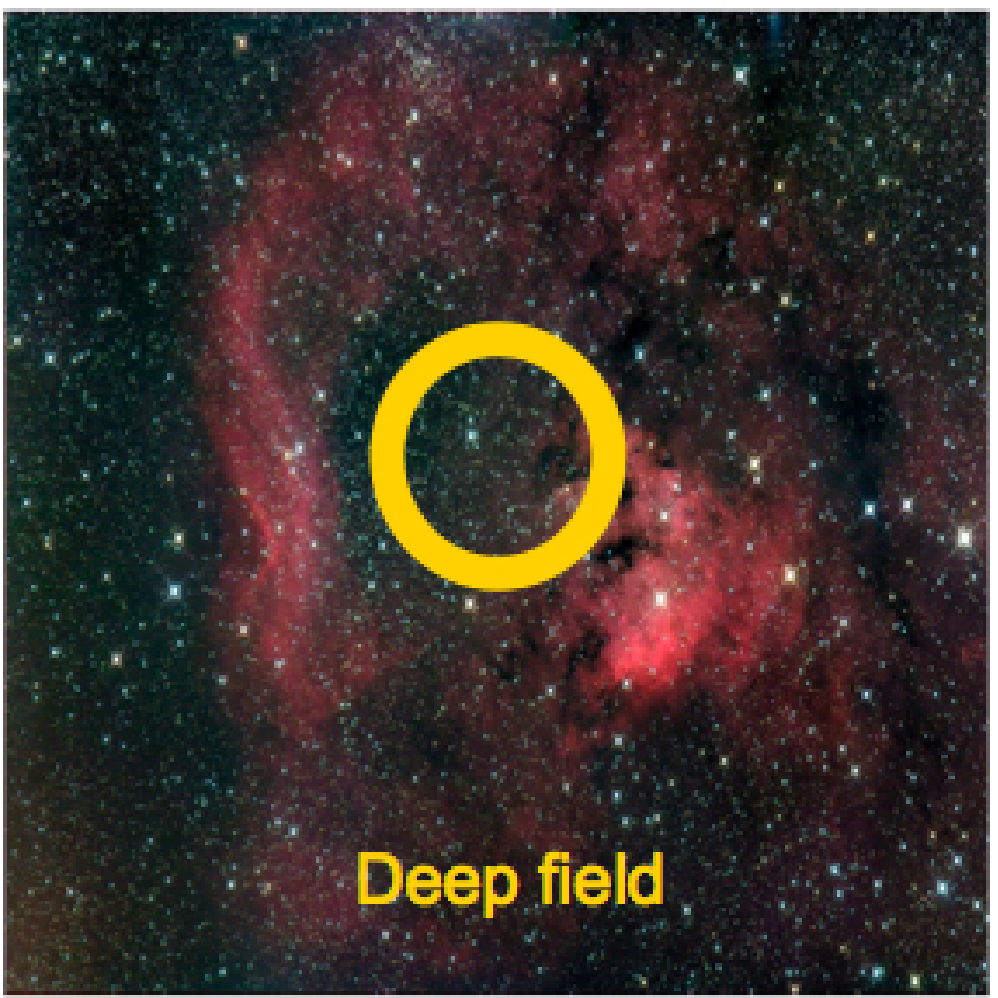}
\includegraphics[width=0.32\linewidth]{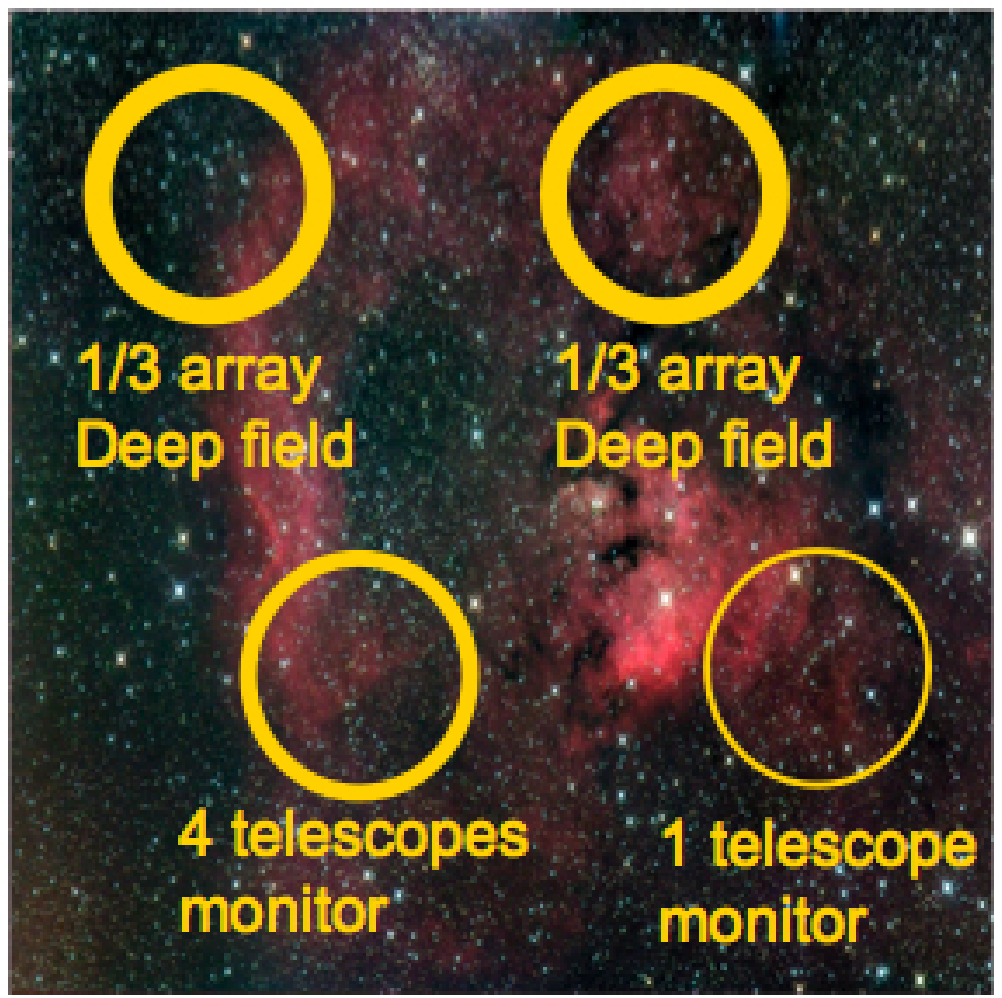}
\includegraphics[width=0.32\linewidth]{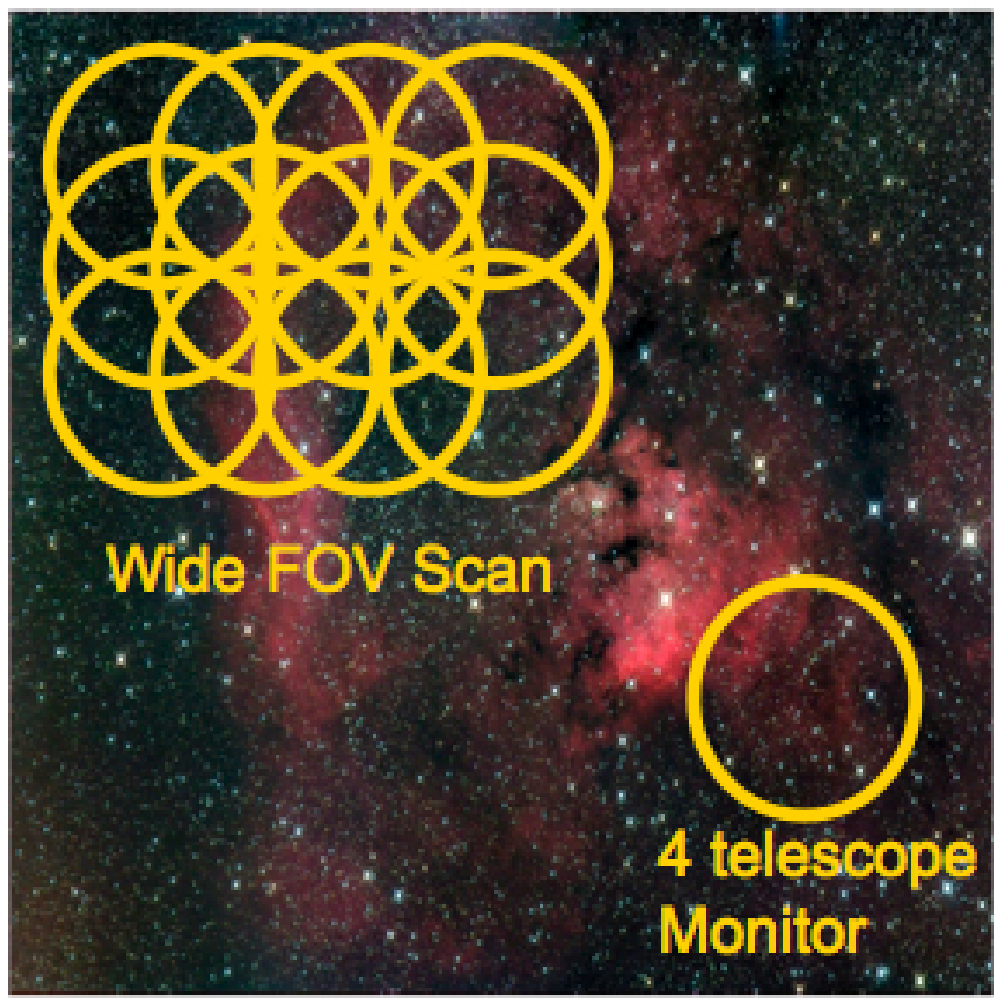}
\caption{\label{fig:cta_modes}Possible operation modes for CTA. From
  left to right: deep field, flexible and wide field modes. More
  details in the text.}
\end{figure}

The trigger systems will be flexible for different operation modes. In
Figure~\ref{fig:cta_modes} some observation modes are shown. In the
deep field mode, all telescopes will be pointed to the same sky
position to maximize the sensitivity. In a more flexible mode, a
fraction or two of the telescopes could point to different positions,
with few telescopes making follow-up observation of single sources as
to monitor blazar activity for example. Finally, telescopes could be
operated in wide-FOV mode, pointed at slightly off-centered region, to
perform an all-sky scan.

\section{The CTA consortium}
The CTA consortium is a partnership between the HESS and MAGIC
collaborations plus several European institutes and recent interest
from world-wide institutions. Activities are coordinated and discussed
with the US AGIS (Advanced Gamma-ray Imaging System) scientists, who
work on a similar project. The consortium hosts already around 50
institutes, 14 countries and 300 scientists. Regular meetings take
place since 2007. The project will be run as observatory. 

The CTA consortium has organized in several working packages (WPs)
each performing research on separated topics. There are 12 WPs: MGT, PHYS, MC,
SITE, MIR, TEL, FPI, ELEC, ATAC, OBS, DATA, QA (standing for
Management, Physics,
Monte Carlo, Site, Mirror, Telescope, Focal-Plane instrumentation,
Electronics, Atmospheric Transmission and Calibration, Observatory,
Data, and Quality Assets).  

A tentative schedule of activities (funds dependent) is reported in
Table~\ref{tab:schedule}. 

\begin{table}[h!]
\centering
\footnotesize{%
\begin{tabular}{p{2.5cm}|ccccccc}
\hline
& '09 & '10 & '11 & '12 & '13 & '14 & '15\\
\hline
Site exploration  &
\multicolumn{1}{>{\columncolor[gray]{0.6}}c}{} &
\multicolumn{1}{>{\columncolor[gray]{0.6}}c}{} &
\multicolumn{1}{>{\columncolor[gray]{0.8}}c}{} &
&&&\\
Array layout      &
\multicolumn{1}{>{\columncolor[gray]{0.6}}c}{} &
\multicolumn{1}{>{\columncolor[gray]{0.6}}c}{} &
\multicolumn{1}{>{\columncolor[gray]{0.6}}c}{} &
\multicolumn{1}{>{\columncolor[gray]{0.8}}c}{} &
&&\\
Telescope design  &
\multicolumn{1}{>{\columncolor[gray]{0.6}}c}{} &
\multicolumn{1}{>{\columncolor[gray]{0.6}}c}{} &
\multicolumn{1}{>{\columncolor[gray]{0.6}}c}{} &
\multicolumn{1}{>{\columncolor[gray]{0.8}}c}{} &
&&\\
Component protot. & &
\multicolumn{1}{>{\columncolor[gray]{0.6}}c}{} &
\multicolumn{1}{>{\columncolor[gray]{0.6}}c}{} &
\multicolumn{1}{>{\columncolor[gray]{0.6}}c}{} &
\multicolumn{1}{>{\columncolor[gray]{0.8}}c}{} &
&\\
Array prototype   &&&&
\multicolumn{1}{>{\columncolor[gray]{0.6}}c}{} &
\multicolumn{1}{>{\columncolor[gray]{0.6}}c}{} &
\multicolumn{1}{>{\columncolor[gray]{0.8}}c}{} &\\
Array construction &&&&&&
\multicolumn{1}{>{\columncolor[gray]{0.6}}c}{} &
\multicolumn{1}{>{\columncolor[gray]{0.6}}c}{} \\
Partial operation &&&&&&
\multicolumn{1}{>{\columncolor[gray]{0.6}}c}{} &
\multicolumn{1}{>{\columncolor[gray]{0.6}}c}{} \\
\hline
\end{tabular}
}
\caption{\label{tab:schedule} CTA tentative schedule of activities.}
\end{table}

\section{Summary}
For a new generation of IACTs, it is mandatory to: extend energy range
from few tens of GeV to 100 TeV, improve sensitivity and energy
resolution, have a larger FOV and better angular resolution, operate
as observatory, perform multi-wavelength observations. In 5+ years
from now, CTA will open the era of precision GR astronomy, with
many new galactic and extragalactic objects found. CTA may answer
long-standing questions on cosmic-rays: where galactic and
extra-galactic CRs are accelerated, how CRs are accelerated
(hadrons/leptons, jets, magnetic irregularities, etc.). In
addition, CTA may answer fundamental physics like DM, Lorentz
invariance, universe transparency, photon-axion oscillation.\\

{\it Acknowledgement.} I would like to thanks M.~Gaug, N.~Godinovic,
B.~Khelifi and M.~Persic for useful comments. 

\end{document}